\documentclass[nofootinbib,twocolumn,showpacs,preprintnumbers,pre,aps,superscriptaddress]{revtex4-1}

\usepackage[dvipdfmx]{graphicx}
\usepackage{bm}
\usepackage{amsmath}
\usepackage{amssymb}
\usepackage{color}

\begin{document}

\title{Emergence of odd elasticity in a microswimmer using deep reinforcement learning}

\author{Li-Shing Lin}
\affiliation{
Department of Chemistry, Graduate School of Science,
Tokyo Metropolitan University, Tokyo 192-0397, Japan}
\affiliation{
Research Institute for Mathematical Sciences, 
Kyoto University, Kyoto 606-8502, Japan}

\author{Kento Yasuda}
\affiliation{
Research Institute for Mathematical Sciences, 
Kyoto University, Kyoto 606-8502, Japan}

\author{Kenta Ishimoto}
\affiliation{
Research Institute for Mathematical Sciences, 
Kyoto University, Kyoto 606-8502, Japan}

\author{Shigeyuki Komura}\email{komura@wiucas.ac.cn}
\affiliation{
Wenzhou Institute, University of Chinese Academy of Sciences, 
Wenzhou, Zhejiang 325001, China} 
\affiliation{
Oujiang Laboratory, Wenzhou, Zhejiang 325000, China}
\affiliation{
Department of Chemistry, Graduate School of Science,
Tokyo Metropolitan University, Tokyo 192-0397, Japan}


\begin{abstract}
We use the Deep Q-Network with reinforcement learning to investigate the emergence of odd elasticity in an elastic microswimmer model. 
For an elastic microswimmer, it is challenging to obtain the optimized dynamics due to the intricate elastohydrodynamic interactions.
However, our machine-trained model adopts a novel transition strategy (the waiting behavior) to optimize the locomotion.
For the trained microswimmers, we evaluate the performance of the cycles by the product of the loop area (called \textit{non-reciprocality}) and the loop frequency, and show that the average swimming velocity is proportional to the performance. By calculating the force-displacement correlations, we obtain the effective odd elasticity of the microswimmer to characterize its non-reciprocal dynamics. This emergent odd elasticity is shown to be closely related to the loop frequency of the cyclic deformation. Our work demonstrates the utility of machine learning in achieving optimal dynamics for elastic microswimmers and introduces post-analysis methods to extract crucial physical quantities such as non-reciprocality and odd elasticity.
\end{abstract}

\maketitle

\section{Introduction}
\label{sec:introduction}

Active systems composed of self-driven units play a crucial role in biological processes as 
they are able to convert microscopic energy into macroscopic 
work~\cite{Marchetti2013HydrodynamicsMatter,Roadmap20,Shanker22}.
To achieve sustainable work at the microscopic scale, active units must go through non-reciprocal 
cyclic motions~\cite{Dey19,HK22}.
For example, cyclic state transitions of enzymatic molecules are driven by catalytic chemical 
reactions~\cite{Dey16,Aviram18}, which can be described by simple coarse-grained 
models~\cite{Echeverria11,Mikhailov15,Hosaka20}.
To evaluate the functionality of an enzyme, we previously defined a physical quantity called 
\textit{non-reciprocality} that represents the area enclosed by a trajectory in the conformational 
space~\cite{Yasuda21-a,Kobayashi23b}.
According to Purcell's scallop theorem for microswimmers moving in a viscous 
fluid~\cite{Purcell77, Lauga09a,Laugabook}, the average swimming velocity is proportional to the non-reciprocality 
and the loop frequency of the cyclic body motion.
It was also reported that the crawling speed of a cell on a substrate is determined by the 
non-reciprocality~\cite{Leoni15,Lenoi17,Tarama18}.

Recently, Scheibner~\textit{et al.}~introduced the concept of odd elasticity which is useful for characterizing 
non-equilibrium active systems~\cite{Scheibner2020OddElasticity_ml,Fruchar2023}. 
Odd elasticity, arising from antisymmetric (odd) components of the elastic modulus tensor that violate the 
energy conservation law, can exist in active materials~\cite{braverman2021,bililign21,surowka2022,Fossati23},  
biological systems~\cite{Tan22}, and active robots~\cite{Ishimoto22_ml,Ishimoto23,Coulais2021}.
We emphasize that the concept of odd elasticity is not limited to elastic materials but can be 
extended to various dynamical systems~\cite{Yasuda22-machlup,Yasuda22}.
An illustrative example is that a microswimmer with odd elasticity can exhibit directional locomotion in the presence of 
thermal agitation~\cite{Yasuda21}.
In fact, the average velocity of an odd microswimmer is proportional to the odd elasticity.
On the other hand, in the model of a stochastic enzyme, we have quantified the average work per
cycle in terms of effective odd elasticity~\cite{Kobayashi23b}. 
Notably, odd elasticity serves as a useful measure for characterizing non-equilibrium micromachines 
such as proteins, enzymes, microswimmers, and robots, regardless of their specific functions.

Despite the importance of odd elasticity in active systems, its physical origin still needs 
to be better understood~\cite{Mitarai02}. 
One possibility is to use Onsager's variational principle~\cite{DoiSoftMatterPhysics_ml} 
to derive dynamical equations for an active system with odd elasticity~\cite{Lin23}.
The obtained non-reciprocal equations~\cite{Fruchart2021} manifest the physical origin of the odd 
elastic constant that is proportional to the non-equilibrium driving force~\cite{Lin23}.
On the other hand, odd elasticity may not be innate to micromachines but can be an ability acquired 
after many experiences and training processes. 
In this work, considering an elastic three-sphere microswimmer 
model~\cite{Pande15,Pande17,Yasuda17c}, we utilize machine learning techniques to account for 
the emergence of odd elastic relation between its elastic components.
With this approach, a microswimmer can automatically obtain the most efficient swimming strategy 
without prescribing any deformation dynamics.

In recent years, machine learning has been widely applied to active systems as a powerful tool to unravel 
the complexities of biological systems~\cite{cichos2020,stark2021,hartl2021,stark2023}. 
Notably, the application of reinforcement learning techniques is capable and versatile in training 
various microswimmers.
These methods have been used to navigate them through complex and dynamic environments with 
remarkable adaptability, such as path-planning in turbulent flows or 
noisy surroundings~\cite{schneider2019,alageshan2020,landin2021,nasiri2023}. 
Furthermore, machine learning has been applied to optimize the local motion and intricate navigation of 
microswimmers with complex structures or higher degrees of freedom~\cite{tsang2020, zou2022, qin2023}. 
These approaches have been further extended to more difficult tasks, such as cooperative swimming and 
predation models~\cite{borra2022,zhu2022,liu2023}.

The main aim of this article is to reveal the emergence of odd elasticity in an elastic microswimmer model by using the Deep Q-Network with reinforcement learning. Traditional kinematic models, such as rigidly connected swimmers, select paths in the deformation space (gait-switching) assuming that the applied forces can instantaneously adapt to any prescribed motion~\cite{tsang2020, Golestanian2004, Golestanian2008}. However, the deformation of an elastic microswimmer cannot be prescribed~\cite{Yasuda17c}. Determining these dynamics requires a nuanced understanding of the elastohydrodynamic process, where traditional methods for achieving optimal control face significant challenges. These challenges are evident in the previous study~\cite{Yasuda17c}, in which the swimming velocity of an elastic microswimmer decreases in its large-frequency regime when prescribed dynamics are assumed.

Unlike the prescribed dynamics model, our machine-learning approach successfully develops an optimal control strategy, adapting a transition (emergence of the waiting behavior) to avoid the velocity decrease. We note that such elastohydrodynamic systems, different from the study in Tsang~\textit{et al.}~\cite{tsang2020}, usually require continuous state and action spaces to tackle. The newly discovered strategy transition using the waiting behavior emerges from the fluid-structure interactions in which distinct hydrodynamic modes with different time scales play important roles.

From the obtained numerical data, we quantify the performance and effective odd elasticity of 
the trained microswimmer. 
The estimated cycle performance, which is the product of the non-reciprocality (loop area) and 
the loop frequency, coincides with the swimming velocity by using a proper scale factor.  
We also demonstrate that the emergent odd elasticity of the microswimmer is closely related to the 
loop frequency of the cyclic deformation.
The present work demonstrates the utility of machine learning in revealing various non-reciprocal 
phenomena in active systems.

In Sec.~\ref{sec:elasticswimmer}, we review the model of an elastic microswimmer with prescribed dynamics~\cite{Yasuda17c}.  
In Sec.~\ref{sec:DQN}, we explain the deep reinforcement learning technique to train the elastic 
microswimmer. 
In Sec.~\ref{sec:trained}, we describe the physical properties of the fully trained microswimmer.
In particular, we shall discuss the emergence of limit cycles, cycle performance, average velocity, and 
effective odd elasticity. 
In Sec.~\ref{sec:training}, we briefly mention the training progression of an elastic microswimmer.
A summary of our work and some discussion are given in Sec.~\ref{sec:summary}.

\begin{figure}[tb]
\centering
\includegraphics[scale=0.23]{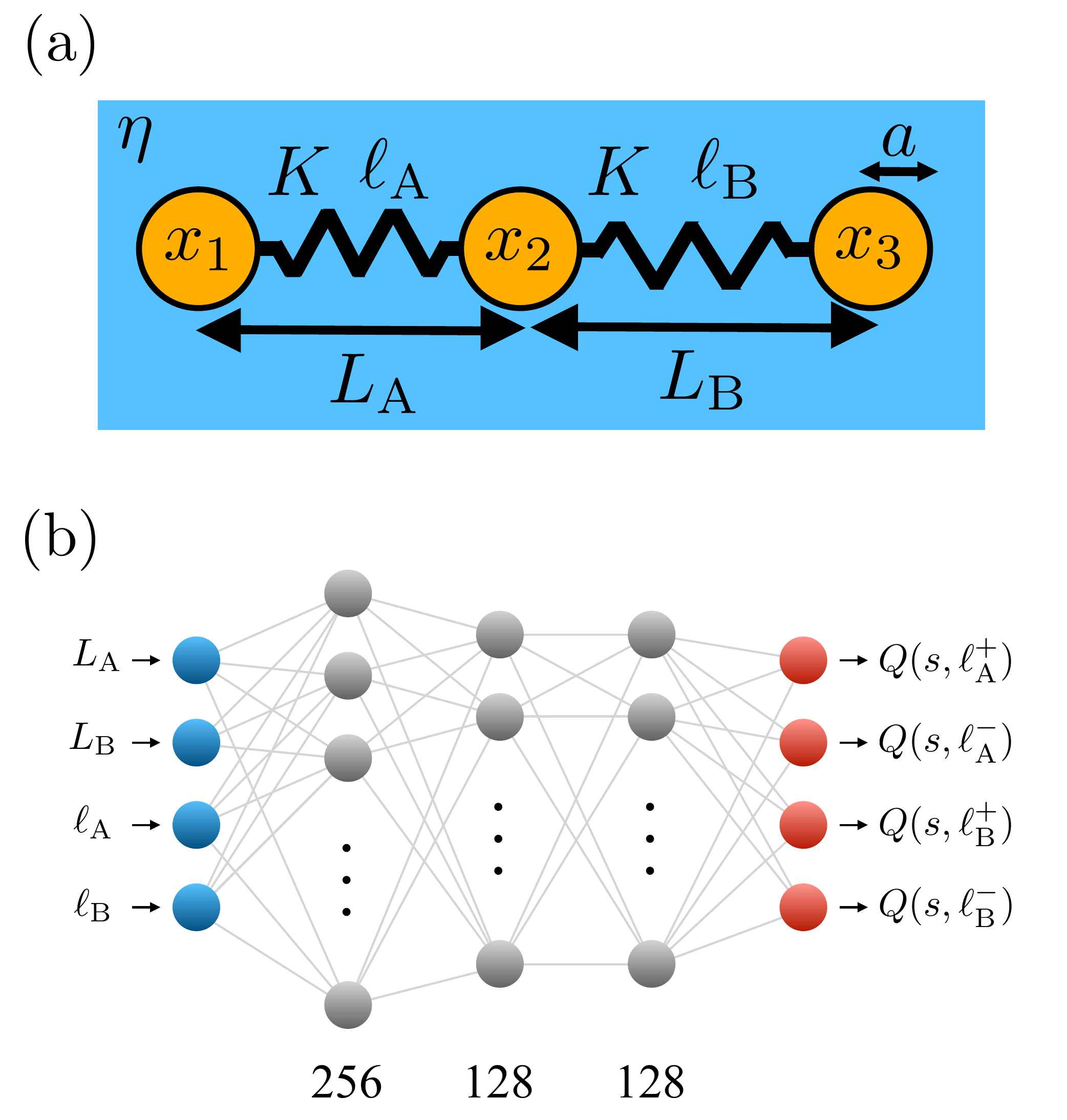}
\caption{(a) 
An elastic microswimmer consists of three spheres of radius $a$ positioned along a one-dimensional 
coordinate system, denoted by $x_i$ ($i=1,2,3$). 
The three spheres are connected by two harmonic springs with the elastic constant $K$ and the 
time-dependent natural lengths $\ell_{\alpha}$ ($\alpha = \rm{A, B}$).
The spring extensions are denoted by $L_{\rm A} = x_2 - x_1$ and $L_{\rm B} =  x_3 - x_2$.
The microswimmer is immersed in a viscous fluid with the shear viscosity $\eta$ and $\tau = 6\pi\eta a/K$ 
gives the hydrodynamic relaxation time. 
(b) Schematic of the neural network architecture that predicts Q-values for each action from an input observation
of $L_{\alpha}$ and $\ell_{\alpha}$.
The blue and red columns represent the input and output layers, respectively. 
The grey columns are the three linear hidden layers with dimensions 256, 128, and 128.
}
\label{elastic_network}
\end{figure}

\section{Elastic microswimmer with prescribed dynamics}
\label{sec:elasticswimmer}

\begin{figure}[tb]
\centering
\includegraphics[scale=0.5]{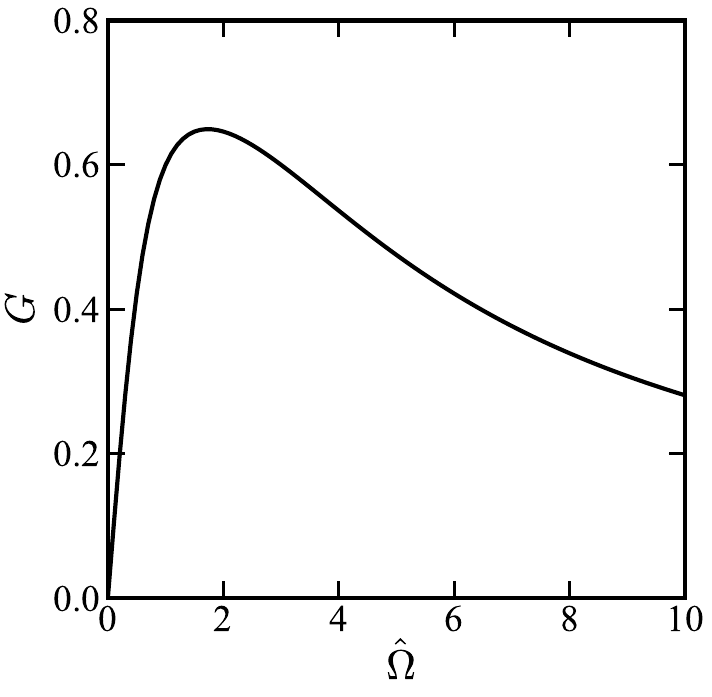}
\caption{
The scaling function $G(\hat{\Omega})$ in Eq.~(\ref{scaling}) that describes the frequency 
dependence of the average swimming velocity $\overline{V}$ of an elastic swimmer 
with prescribed dynamics of the natural lengths given by Eqs.~(\ref{prescribed_la}) and 
(\ref{prescribed_lb}).
}
\label{pre_dependency}
\end{figure}

We first review the elastic three-sphere microswimmer model introduced by the present 
authors~\cite{Yasuda17c,Kuroda19,Era21,Yasuda2023} and others~\cite{Pande15, Pande17, Grosjean2016}. 
As illustrated in Fig.~\ref{elastic_network}(a), the microswimmer consists of three spheres of radius 
$a$ positioned along a one-dimensional coordinate system, denoted by $x_i$ ($i=1,2,3$). 
One can assume $x_1< x_2 < x_3$ without loss of generality. 
Unlike the original three-sphere microswimmer model by Najafi and 
Golestanian~\cite{Golestanian2004,Golestanian2008}, 
the three spheres are connected by two harmonic springs, each with a time-dependent 
natural length $\ell_{\alpha}$ ($\alpha = \rm{A, B}$) and having the same spring elastic constant $K$.
Such an \textit{elastic microswimmer} is immersed in a viscous fluid with the shear viscosity $\eta$. 
Although $a$ and $K$ can differ between the spheres and the springs, 
respectively~\cite{Kobayashi23a,Yasuda17c,Golestanian2008}, we consider here the symmetric case. 
This elastic microswimmer model~\cite{Yasuda17c} reduces to the original three-sphere model with 
rigid arms~\cite{Golestanian2004,Golestanian2008} when $K$ is infinitely large.

When the spring lengths $L_{\rm A} = x_2 - x_1$ and $L_{\rm B} =  x_3 - x_2$ deviate from their natural 
lengths $\ell_{\alpha}$, the elastic forces $f_i$ acting on each sphere are given by
\begin{align}
 f_1 & = K(L_{\rm A} - \ell_{\rm A}), 
 \label{f1}
 \\ 
f_2 & = -K(L_{\rm A} - \ell_{\rm A}) + K(L_{\rm B} - \ell_{\rm B}), 
\label{f2}
\\
f_3 & = -K(L_{\rm B} - \ell_{\rm B}). 
\label{f3}
\end{align}
Due to hydrodynamic interactions described by the Stokes mobility and the Oseen tensor, 
the forces $f_i$ and the sphere velocities $v_i=\dot{x}_i = dx_i/dt$ (dot indicates the time 
derivative) are related by~\cite{Golestanian2004,Golestanian2008}
\begin{align}
v_1 &= \frac{f_1}{6 \pi \eta a} + \frac{f_2}{4 \pi \eta L_{\rm A}} + \frac{f_3}{4 \pi \eta (L_{\rm A} + L_{\rm B})}, 
\label{v1}
\\
v_2 &= \frac{f_1}{4 \pi \eta L_{\rm A}} + \frac{f_2}{6 \pi \eta a} + \frac{f_3}{4 \pi \eta L_{\rm B}}, 
\label{v2}
\\
v_3 &= \frac{f_1}{4 \pi \eta (L_{\rm A} + L_{\rm B})} + \frac{f_2}{4 \pi \eta L_{\rm B}} + \frac{f_3}{6 \pi \eta a}, 
\label{v3}
\end{align}
where the conditions $a/L_\alpha \ll 1$ are assumed.
For the elastic microswimmer, the force-free condition, $f_1 + f_2 + f_3 = 0$, is automatically satisfied,
ensuring a self-propelled motion without any external force.

When $L_{\rm \alpha}$ and $\ell_{\alpha}$ are given, the dynamics of the microswimmer are deterministic 
and the total swimming velocity $V = (v_1 + v_2 + v_3)/3$ is given by 
\begin{align}
V = & \frac{1}{12\pi\eta} \biggl[\left(\frac{1}{L_{\rm A} + L_{\rm B}} - \frac{1}{L_{\rm B}} \right) f_1 
\nonumber
\\
& +\left(\frac{1}{L_{\rm A} + L_{\rm B}} - \frac{1}{L_{\rm A}}\right) f_3\biggr]. 
\label{CMV}
\end{align}
For relatively small deformations of the springs, we can define the small displacements with 
respect to the average spring length $\ell$ as $u_{\alpha} = L_\alpha - \ell$ ($\alpha = \rm{A, B}$).
Within the small-amplitude approximation, $u_{\alpha}/\ell \ll 1$, Golestanian and Ajdari calculated the 
average swimming velocity of a three-sphere microswimmer up to the leading order 
as~\cite{Golestanian2008}  
\begin{align}
\overline{V}=\frac{7a}{24\ell^2}
\overline{(u_{\rm A}\dot{u}_{\rm B}-\dot{u}_{\rm A} u_{\rm B})}.
\label{general_average_speed}
\end{align}
Here the averaging, indicated by the bar, is performed by time integration in a full cycle and 
further divided by the total time of a period. 
The above expression indicates that $\overline{V}$ is determined by the product of the closed 
loop area and the loop frequency~\cite{Shapere89}.

The explicit form of $\overline{V}$ of an elastic microswimmer can be obtained by specifying a prescribed 
cyclic change in the natural spring lengths $\ell_{\alpha}$. 
Previously, we used the following sinusoidal forms~\cite{Yasuda17c,Yasuda2023} 
\begin{align}
\ell_{\rm A} & = \ell + d_{\rm A} \cos(\Omega t),
\label{prescribed_la}
\\
\ell_{\rm B} & = \ell + d_{\rm B} \cos(\Omega t - \phi),
\label{prescribed_lb}
\end{align}
where $\ell$ is the constant natural length, $d_{\alpha}$ are the amplitudes of the 
oscillatory change, $\Omega$ is the common frequency, and $\phi$ is the phase difference between the 
two cyclic changes.
When the natural lengths undergo this prescribed cycle, the spring lengths relax to their new natural lengths 
obeying Eqs.~(\ref{f1})-(\ref{v3}) with a hydrodynamic relaxation time  
$\tau = 6 \pi \eta a/K$.

Then the average swimming velocity of an elastic microswimmer with the prescribed dynamics 
was calculated to be~\cite{Yasuda17c,Yasuda2023}
\begin{align}\overline{V} = \frac{7ad_{\rm A}d_{\rm B}}{24 \ell^2 \tau} G(\hat{\Omega}) \sin \phi,
\label{prescribed_swimmer_average}
\end{align}
where $\hat{\Omega}=\Omega \tau$ is the dimensionless frequency and $G(\hat{\Omega})$ is the scaling function
\begin{align}
G(\hat{\Omega})=\frac{3\hat{\Omega} (3+\hat{\Omega}^2)}{9+10\hat{\Omega}^2+\hat{\Omega}^4}.
\label{scaling}
\end{align}
We see that $\overline{V}$ is non-zero when $\phi \neq 0$, corresponding to the non-reciprocal deformation, 
and $\vert \overline{V} \vert$ is maximized when $\phi=\pm \pi/2$. 
Notably, such non-reciprocal deformation can be effectively generated by assuming an antisymmetric stress-strain 
relation among the two springs~\cite{Yasuda22, Yasuda21}. 
This antisymmetric cross-correlation represents what we refer to as odd elasticity in this article.

In Fig.~\ref{pre_dependency}, we plot $G(\hat{\Omega})$ using Eq.~(\ref{scaling}).
In the small-frequency limit, where $\hat{\Omega} \ll 1$, the average velocity increases as 
$\overline{V} \sim \hat{\Omega}$~\cite{Golestanian2004,Golestanian2008}.
However, in the larger-frequency limit, where $\hat{\Omega} \gg 1$, the average velocity decreases as 
$\overline{V} \sim \hat{\Omega}^{-1}$~\cite{Yasuda17c,Yasuda2023}.
The crossover frequency between these two regimes is approximately $\hat{\Omega} \approx 1$.
In the large-frequency regime, the mechanical response is delayed because it takes time for the springs to 
relax to their natural lengths.
Such a decrease in the swimming velocity is a drawback of elastic microswimmers with prescribed motion.
A similar crossover behavior and a decrease in the average velocity were also predicted for the 
Najafi-Golestanian microswimmer model in a viscoelastic fluid~\cite{Yasuda2017,YasudaKuroda20}.

\begin{figure*}[tb]
\centering
\includegraphics[scale=0.5]{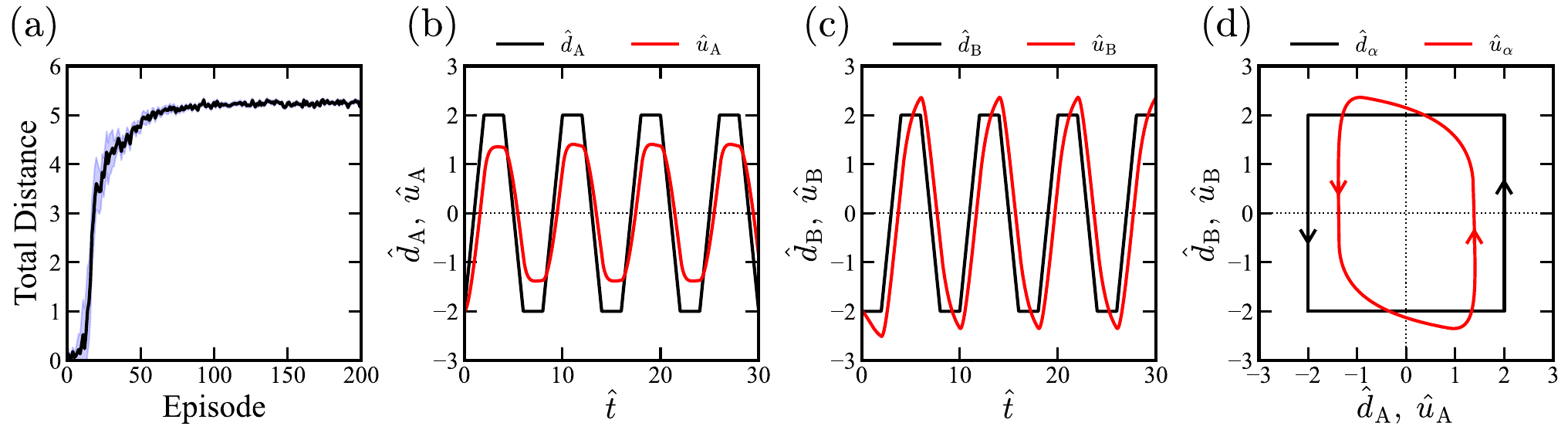}
\caption{
(a) Training curve of an elastic microswimmer with an actuation velocity of $\hat{U} = 2$. 
The black line represents the total distance achieved by the microswimmer in each training episode for a single case. 
The blue shaded area indicates the standard deviation around the mean value obtained from 100 training 
instances of microswimmers with the same actuation velocity. 
(b), (c) Periodic dynamics of the natural length deviations 
$\hat{d}_\alpha = \hat{\ell}_{\alpha}-\hat{\ell}$ (black) and the the spring extensions 
$\hat{u}_{\alpha}=\hat{L}_\alpha - \hat{\ell}$ (red) obtained from the fully trained microswimmer with 200 
episodes of experience.
The dimensionless time is defined by $\hat{t}=t/\tau$, and the parameters are $\hat{U} = 2$ and $\hat{\ell}=10$.
The oscillation phase difference between (b) and (c) is approximately $\pi/2$, with a fore-aft amplitude difference.
(d) The configuration space trajectory of $\hat{d}_\alpha$ (black) and $\hat{u}_{\alpha}$ (red)
when $\hat{U} = 2$. 
Both $\hat{d}_\alpha$ and $\hat{u}_{\alpha}$ form counterclockwise cyclic loops. 
}
\label{u2_plot}
\end{figure*}

\section{Elastic microswimmer directed by reinforcement learning}
\label{sec:DQN}

Using an elastic microswimmer model, we apply a machine learning method to direct its movement rather 
than prescribing the motion of the natural lengths $\ell_{\alpha}$.
We combine the Deep Q-Network (DQN) with reinforcement learning~\cite{Dayan1992, mnih2013, sutton2018} 
to train the actuation of a microswimmer and obtain the optimized dynamics for the natural lengths.
In particular, we shall investigate how a trained microswimmer adapts to a new strategy to avoid the 
decrease in the average swimming velocity when the actuation is faster than the hydrodynamic relaxation.

As shown in Fig.~\ref{elastic_network}(b),
our artificial intelligence (AI) uses the spring lengths $L_{\rm \alpha}$ and the natural lengths 
$\ell_{\rm \alpha}$ as an observation (state) input and performs an action to change either 
$\ell_{\rm A}$ or $\ell_{\rm B}$ with an \textit{actuation velocity} $U$, i.e., the rate of changes in the 
natural lengths. 
Specifically, the output action space is discrete and consists of four actions $\ell_{\alpha}^\pm$ 
corresponding to the increase and decrease of $\ell_{\alpha}$.
When the natural lengths $\ell_\alpha$ change, the spring lengths $L_\alpha$ tend to relax 
toward the new natural lengths according to Eqs.~(\ref{f1})-(\ref{v3}).  
Hereinafter, we choose the sphere radius $a$ and the hydrodynamic relaxation time 
$\tau= 6 \pi \eta a/K$ as the units for length and time, respectively.
The dimensionless quantities are then denoted with a hat such as  
$\hat{L}_{\alpha} = L_\alpha/a$, $\hat{\ell}_{\alpha} = \ell_\alpha/a$, and $\hat{U}=U \tau/a$.

During the training process, we constrain the natural lengths in the range $8 \le \hat{\ell}_\alpha \le 12$ 
to ensure the conditions $a/L_\alpha \ll 1$ and $(L_\alpha-\ell)/\ell\ll1$. 
This assignment simplifies the model and aligns with established methods, enabling valid comparisons 
with existing studies~\cite{Yasuda21, Yasuda2023, Golestanian2004, Golestanian2008}. 
In our model, the actuation velocity $\hat{U}$ is a control parameter and can be contrasted
with the frequency $\hat{\Omega}$ in Sec.~\ref{sec:elasticswimmer}.
We train the microswimmers at different actuation velocities ranging from $\hat{U} = 0.1$ to $10$, 
with an increment of $0.1$.
Each training session comprises 200 episodes, and each episode contains 1,200 decisions made by the AI.

To optimize the locomotion of the microswimmer, we initiate each episode with a random state and employ 
the Epsilon Greedy Algorithm (EGA) to further balance between exploration and exploitation during training
process~\cite{sutton2018}.
Actions are determined successively by the AI in each decision step in which the natural lengths 
$\hat{\ell}_{\alpha}$ are changed by $\pm 1$ with a given actuation velocity $\hat{U}$. 
Hence, consecutive decisions are made at intervals of $\Delta \hat{t}_{\rm dec} = 1/\hat{U}$. 
Since the numerical time step $\Delta \hat{t}_{\rm{num}} = 0.01$ is used to solve the 
hydrodynamic equations in Eqs.~(\ref{f1})-(\ref{v3}), the relation 
$\Delta \hat{t}_{\rm dec}/ \Delta \hat{t}_{\rm num}= 100/\hat{U}$ ensures adequate 
numerical time steps between successive actions for $0.1 \le \hat{U} \le 10$.

In our model, successive actions are consistently taken at intervals of $\Delta \hat{t}_{\rm dec}$. 
This means that the subsequent action is executed immediately after $\hat{\ell}_\alpha$ has evolved 
by $\pm 1$, irrespective of whether the spring lengths $\hat{L}_{\alpha}$ have fully relaxed to the 
new natural lengths $\ell_{\alpha}$ or not. 
In specific situations, the machine can conduct a waiting strategy where it refrains from changing any 
natural lengths for $\Delta \hat{t}_{\rm{dec}}$. 
Such a decision arises from actions that violate the natural length constraints. 
For example, taking $\hat{\ell}_{\rm A}^+$ is not allowed when $\hat{\ell}_{\rm A}=12$.
In this case, no change is applied to $\hat{\ell}_{\rm A}$, and the consecutive action is conducted 
after $\Delta \hat{t}_{\rm dec}$.

The training process is formulated as a Markov decision process (MDP) with the memoryless 
property~\cite{sutton2018}, ensuring that the immediate reward depends only on the current state 
and action.
Within the MDP method, the training of the DQN is guided by Bellman's equation that is used to 
iteratively update the prediction of the Q-value function (the expected cumulative future reward) 
in every decision step~\cite{Dayan1992,mnih2013,sutton2018}.
In our model, Bellman's equation is given by
\begin{align}
Q(s_t, a_t) = r_t + \gamma \max_{a'} Q(s_{t+1}, a'),
\end{align}
where $Q(s_t, a_t)$ represents the Q-value of taking an action $a_t \in \hat{\ell}_{\alpha}^\pm$ 
in state $s_t= (\hat{L}_{\alpha}, \hat{\ell}_{\alpha})$ at time $t$, and $r_t$ denotes the reward obtained after 
taking action $a_t$ in state $s_t$.
The reward is defined as the positive displacement of the whole microswimmer during a decision step. 
The coefficient $\gamma$ represents the discount factor that balances the importance between 
immediate rewards and future rewards. 
We choose a conventional value of $\gamma=0.99$ for farsightedness to prioritize long-term cumulative 
rewards~\cite{tsang2020, zou2022, qin2023, liu2023}. 
By updating the network with Bellman's equation in every decision step, our DQN efficiently learns the 
optimal policy for controlling the dynamics of the natural lengths $\hat{\ell}_{\alpha}$.

In Fig.~\ref{u2_plot}(a), we show the training curve of a microswimmer when the actuation velocity  
$\hat{U} = 2$.
As a function of the trained episodes, we plot the total distance, namely, the net displacement from the initial 
position that the microswimmer can achieve within each episode. 
During the training process, the total distance continuously increases over the episodes, indicating enhanced 
locomotion ability.
After about 100 episodes of training, the microswimmer's performance approaches an optimal swimming 
distance within each single episode.
The small fluctuations that remain after reaching the optimal total distance are due to random initial conditions 
and EGA used for the training. 
Details of the training progression will be discussed separately in Sec.~\ref{sec:training}.

The characteristics of the training curve vary for swimmers with different values of $\hat{U}$. 
As $\hat{U}$ increases, swimmers require more training episodes to achieve optimized swimming velocity. 
This trend becomes pronounced at higher values of $\hat{U}$, which are not included in this article due 
to the lack of additional physical interest (e.g., $\hat{U} \sim $ $20$--$30$). 
On the other hand, the standard deviation of the training datasets shows increased sensitivity within the range 
$\hat{U} \in [0.1, 10]$. 
This increase in standard deviation is due to the complexity arising from the delayed response of the exact length 
$L_\alpha$. 
Such a delay in response leads to a behavior transition, which will be further discussed in Sec.~\ref{sec:trained}.

\section{Fully trained elastic microswimmer}
\label{sec:trained}

In this section, we discuss the dynamic properties of the microswimmers which have been trained
for 200 episodes.
Besides the emergence of limit cycles, we shall discuss the performance and effective odd elasticity 
of trained microswimmers.

\subsection{Emergence of limit cycles}

\begin{figure*}[tb]
\centering
\includegraphics[scale=0.5]{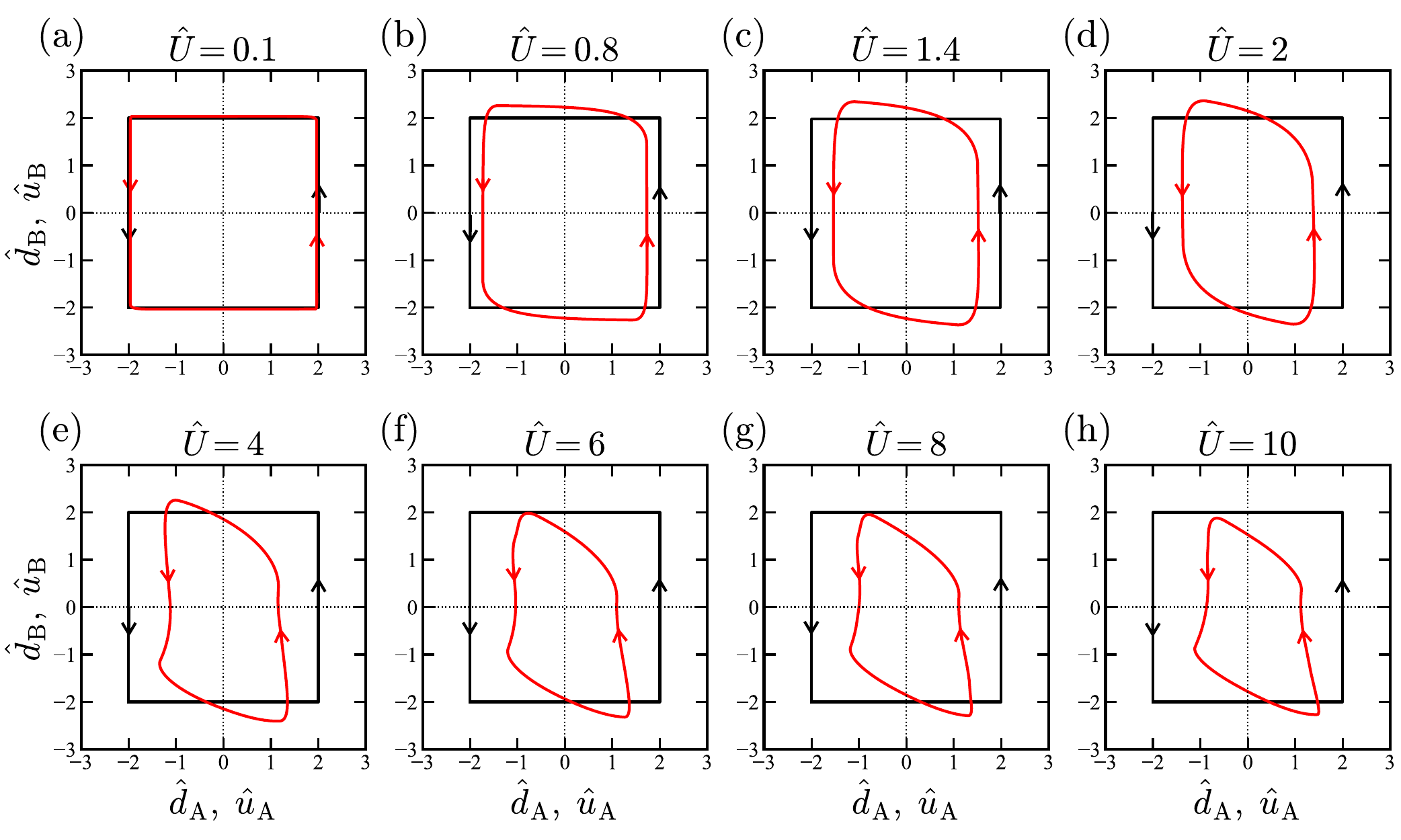}
\caption{
Limit cycles of microswimmers trained with different $\hat{U}$-values ranging from $0.1$ to $10$.
Black and red lines represent the natural length deviations $\hat{d}_\alpha$ and the 
spring extensions $\hat{u}_{\alpha}$, respectively. 
The plot in (d) for $\hat{U}=2$ is identical to that in Fig.~\ref{u2_plot}(d).
}
\label{ex_u0_to_u10}
\end{figure*}

\begin{figure}[tb]
\centering
\includegraphics[scale=0.45]{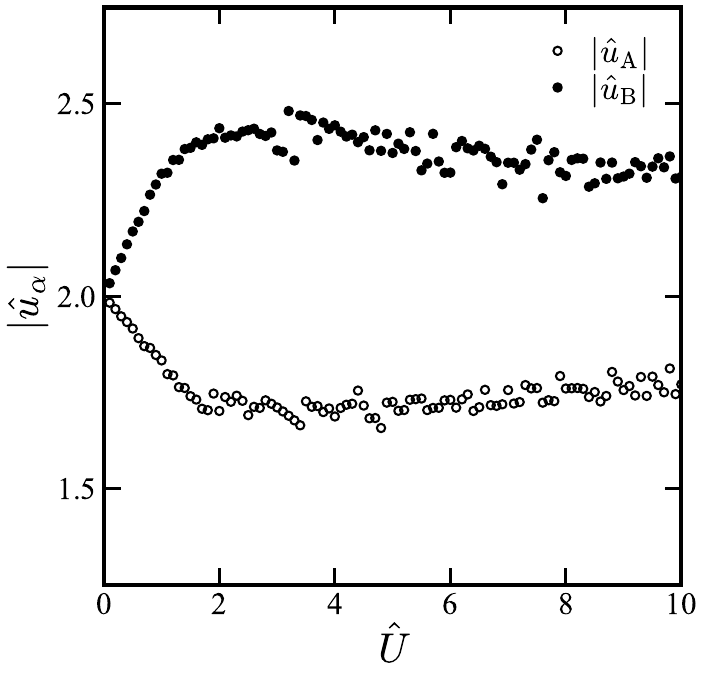}
\caption{
The dimensionless spring amplitudes $\vert \hat{u}_\alpha \vert$ as a function of $\hat{U}$.
The amplitudes exhibit fore-aft asymmetry.
}
\label{amplitude}
\end{figure}

\begin{figure*}[tb]
\centering
\includegraphics[scale=0.5]{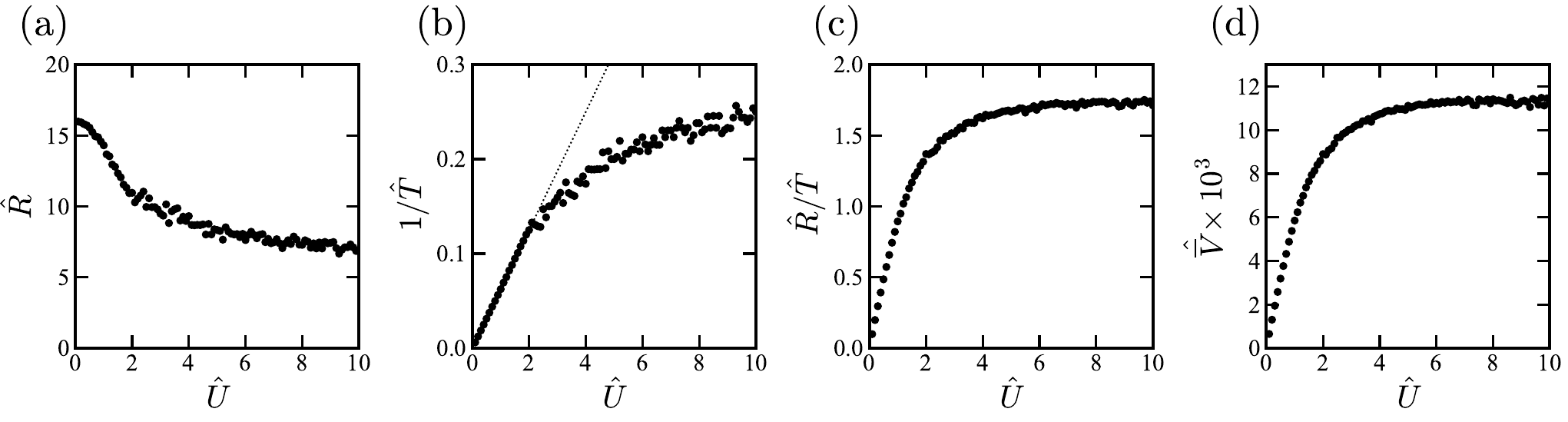}
\caption{
Performance of the fully trained microswimmer at various actuation velocities $\hat{U}$.
(a) The dimensionless non-reciprocality $\hat{R}=R/a^2$ defined in Eq.~(\ref{nonreciprocality}) as a 
function of $\hat{U}$. 
This quantity represents the enclosed area of the red loops made of $u_{\alpha}$ in Fig.~\ref{ex_u0_to_u10}. 
(b) The dimensionless loop frequency $1/\hat{T}=\tau/T$ as a function of $\hat{U}$.
When $\hat{U} \leq 2$, the loop frequency increases linearly with $\hat{U}$ as fitted by the 
dotted line ($1/\hat{T}=\hat{U}/16$).
(c) The dimensionless cycle performance $\hat{R}/\hat{T}$ (calculated from (a) and (b)) as a function 
of $\hat{U}$.
(d) The dimensionless average swimming velocity $\hat{\overline{V}} = \overline{V}\tau/a$ as a function 
of $\hat{U}$.
Both $\hat{R}/\hat{T}$ in (c) and $\hat{\overline{V}}$ in (d) show the same $\hat{U}$-dependence except 
for a scaling factor.
The ratio between these quantities is $\hat{\overline{V}}\hat{T}/\hat{R} \approx 6.57 \times 10^{-3}$.
The interpretation of this ratio is discussed in the text.
}
\label{performance_plot}
\end{figure*}

In Figs.~\ref{u2_plot}(b) and (c), we present the cyclic motions of the fully trained swimmer when $\hat{U} = 2$. 
The black lines represent the deviation of the natural lengths from its average value, i.e.,  
$\hat{d}_\alpha = \hat{\ell}_\alpha - \hat{\ell}$ ($\alpha = \rm{A, B}$), where we chose $\hat{\ell} = 10$
because of the constraint $8 \le \hat{\ell}_\alpha \le 12$.
The red lines represent the spring extensions $\hat{u}_{\alpha}=\hat{L}_\alpha - \hat{\ell}$ ($\alpha = \rm{A, B}$).
Both $\hat{u}_{\rm A}$ and $\hat{u}_{\rm B}$ exhibit periodic 
motions with a phase difference of approximately $\pi/2$, which corresponds to the maximum efficiency for swimming.

In Fig.~\ref{u2_plot}(d), we present the configuration space trajectory of the same trained microswimmer
over one cycle.
The trajectory of the natural lengths $\hat{d}_{\alpha}$ (shown in black) forms a counterclockwise
square in the range $-2 \le \hat{d}_{\alpha} \le 2$. 
The spring extensions $\hat{u}_{\alpha}$ (shown in red) also exhibit a limit cycle in the configuration space.
This indicates that the fully trained elastic microswimmer has acquired steady non-reciprocal spring motion after training.

To investigate the dependence on the actuation velocity $\hat{U}$, we plot  
the configuration space trajectories for different $\hat{U}$-values ranging from $0.1$ to $10$ in 
Figs.~\ref{ex_u0_to_u10}(a)-(h) (Fig.~\ref{ex_u0_to_u10}(d) and Fig.~\ref{u2_plot}(d) are the same).
In all the cases, the natural lengths $\hat{d}_{\alpha}$ (shown in black) follow the same 
counterclockwise square shape. 
When the actuation velocity is small, such as $\hat{U} \le 2$ in Figs.~\ref{ex_u0_to_u10}(a)-(d),
the hydrodynamic relaxation process can catch up with the change in the natural lengths, and the cycles 
of the spring extensions $\hat{u}_{\alpha}$ (shown in red) are close to the square-shaped trajectory. 
In this regime, the enclosed area within the loop decreases when $\hat{U}$ is increased, 
as we quantify in the next subsection.

For larger actuation velocity $\hat{U} > 2$, corresponding to Figs.~\ref{ex_u0_to_u10}(e)-(h), the 
hydrodynamic relaxation becomes the slower mode. 
This situation is similar to the case of $\hat{\Omega} > 1$ in Sec.~\ref{sec:elasticswimmer}. 
For the trained microswimmer, however, the AI adapts a waiting strategy once the natural lengths reach 
the maximum or minimum values ($\hat{d}_{\alpha}=\pm2$) for the spring extensions 
$\hat{u}_{\alpha}$ to relax sufficiently.
This waiting strategy allows the spring lengths to catch up with the large actuation velocity $\hat{U}$ 
and to prevent the enclosed area from further decreasing.
Since the dynamics are no longer dominated by the actuation velocity $\hat{U}$, the trajectories of 
$\hat{u}_{\alpha}$ become less squared and less symmetric for $\hat{U} > 2$.
As a result, the distinctions between different loop shapes become less pronounced in 
Figs.~\ref{ex_u0_to_u10}(e)-(h).

Another notable result in Fig.~\ref{ex_u0_to_u10} is the fore-aft amplitude asymmetry between 
$\hat{u}_{\rm A}$ and $\hat{u}_{\rm B}$.
In Fig.~\ref{amplitude}, we plot the amplitudes $\vert \hat{u}_{\alpha} \vert$ as a function of $\hat{U}$.
Starting from the value $\vert \hat{u}_{\alpha} \vert =2$, $\vert \hat{u}_{\rm A} \vert $ and 
$\vert \hat{u}_{\rm B} \vert$ decreases and increases, respectively, as $\hat{U}$ is increased. 
Notice that the maximum amplitude can be up to $\vert \hat{u}_{\rm B} \vert/\hat{\ell} \approx 0.25$.
When $\hat{U}>2$, however, $\vert \hat{u}_{\alpha} \vert $ are almost independent of $\hat{U}$.
This asymmetry in fore-aft amplitude is due to the non-reciprocal swimming cycle, which is associated 
with the asymmetric order in which $\ell_\alpha$ changes.

\subsection{Cycle performance and swimming velocity}

Next, we discuss the performance of the acquired cyclic motion and the average swimming velocity of 
the fully trained microswimmers.
Following our work on catalytic enzymes~\cite{Yasuda21-a, Kobayashi23b}, we consider the 
following quantity called \textit{non-reciprocality}
\begin{align}
R = \frac{1}{2}\int_{0}^{T} dt \, (u_{\rm A}\dot{u}_{\rm B} - \dot{u}_{\rm A}u_{\rm B}),
\label{nonreciprocality}
\end{align}
where $T$ is the period of one cycle.
We note again that $R$ represents the area enclosed by the loop trajectory in the configuration space~\cite{Shapere89}. 
Then the dimensionless average velocity $\hat{\overline{V}}=\overline{V} \tau /a$ obtained from  
Eq.~(\ref{general_average_speed}) can be rewritten in terms of the dimensionless 
non-reciprocality $\hat{R}=R/a^2$ as  
\begin{align}
\hat{\overline{V}}=\frac{7}{12\hat{\ell}^2} \frac{\hat{R}}{\hat{T}},
\label{dimensionless_speed}
\end{align}
where $\hat{T} = T/\tau$.
We shall call $1/\hat{T}$ the dimensionless \textit{loop frequency}.
The quantity $\hat{R}/\hat{T}$, representing the area enclosed by the loop per unit time, thus determines the 
swimming velocity when the deformation is small.

In Fig.~\ref{performance_plot}(a), we plot the non-reciprocality $\hat{R}$ as a function 
of $\hat{U}$ for fully trained microswimmers. 
As indicated in Fig.~\ref{ex_u0_to_u10}, the enclosed area $\hat{R}$ decreases as $\hat{U}$ increases. 
In Fig.~\ref{performance_plot}(b), the loop frequency $1/\hat{T}$ is plotted as a function 
of $\hat{U}$. 
When $\hat{U} \leq 2$, the loop frequency increases linearly with $\hat{U}$ (shown by the dotted line)
to minimize the loop period.
For $\hat{U} > 2$, however, the dependence of $1/\hat{T}$ on $\hat{U}$ deviates from the linear relation.
This is because the actuation velocity $\hat{U}$ outpaces the hydrodynamic relaxation rate, and 
the AI adapts the waiting strategy to adjust to the slow hydrodynamic mode.
The slope of the dashed line is $1/16$ ($1/\hat{T}=\hat{U}/16$), and hence the dimensionless waiting time at 
$\hat{d}_\alpha = \pm 2$ is estimated by $\hat{T} - 16/\hat{U}$. 
The finite waiting time appears as a transition at $\hat{U} \approx 2$. 
As plotted in Fig.~\ref{performance_plot}(c), we find that the cycle performance, as measured by $\hat{R}/\hat{T}$, 
increases monotonically with $\hat{U}$ and eventually approaches the value $\hat{R}/\hat{T} \approx 1.75$, 
bounded by the hydrodynamic relaxation process.

To check the validity of Eq.~(\ref{dimensionless_speed}), we plot in Fig.~\ref{performance_plot}(d) 
the average swimming velocity $\hat{\overline{V}}$ obtained from the actual displacement of the microswimmers.
Both $\hat{R}/\hat{T}$ and $\hat{\overline{V}}$ in Figs.~\ref{performance_plot}(c) and (d), respectively,
show almost the same dependence on $\hat{U}$ except for a scaling factor. 
The obtained ratio from Figs.~\ref{performance_plot}(c) and (d) is 
$\hat{\overline{V}}\hat{T}/\hat{R} \approx 6.57\times 10^{-3}$ within the studied $\hat{U}$-range.
This value can be compared with the geometrical pre-factor $7/(12\hat{\ell}^2) \approx 5.83\times 10^{-3}$ 
in Eq.~(\ref{dimensionless_speed}) when $\hat{\ell}=10$.  
A small difference between these pre-factors comes from the assumption $u_{\alpha}/\ell \ll 1$ used in 
Eq.~(\ref{general_average_speed}) or Eq.~(\ref{dimensionless_speed}).
As previously shown in Fig.~\ref{amplitude}, the spring deformations for our trained microswimmers can reach 
$\hat{u}_{\alpha}/\hat{\ell} \approx 0.2$ or larger.
Higher-order contributions need to be incorporated into Eq.~(\ref{dimensionless_speed}) to reproduce the 
average swimming velocity achieved by our swimmer model.

The behavior of $\overline{V}$ in Fig.~\ref{performance_plot}(d) for the trained microswimmer is in sharp 
contrast to that of an elastic microswimmer whose natural spring motions are prescribed. 
In Sec.~\ref{sec:elasticswimmer}, we showed in Eq.~(\ref{prescribed_swimmer_average}) and 
Fig.~\ref{pre_dependency} that the average velocity decreases when $\hat{\Omega} > 1$~\cite{Yasuda17c,Yasuda2023}.
For the fully trained microswimmer, however, an emergent waiting strategy appears when $\hat{U} > 2$, such 
that the swimming velocity does not decrease even at higher actuation velocities $\hat{U}$.
This discovery of a strategy transition exemplifies how machine learning can reveal the intricate optimal dynamics 
of complex microswimmers.

\subsection{Effective even and odd elasticities}

\begin{figure}[tb]
\centering
\includegraphics[scale=0.52]{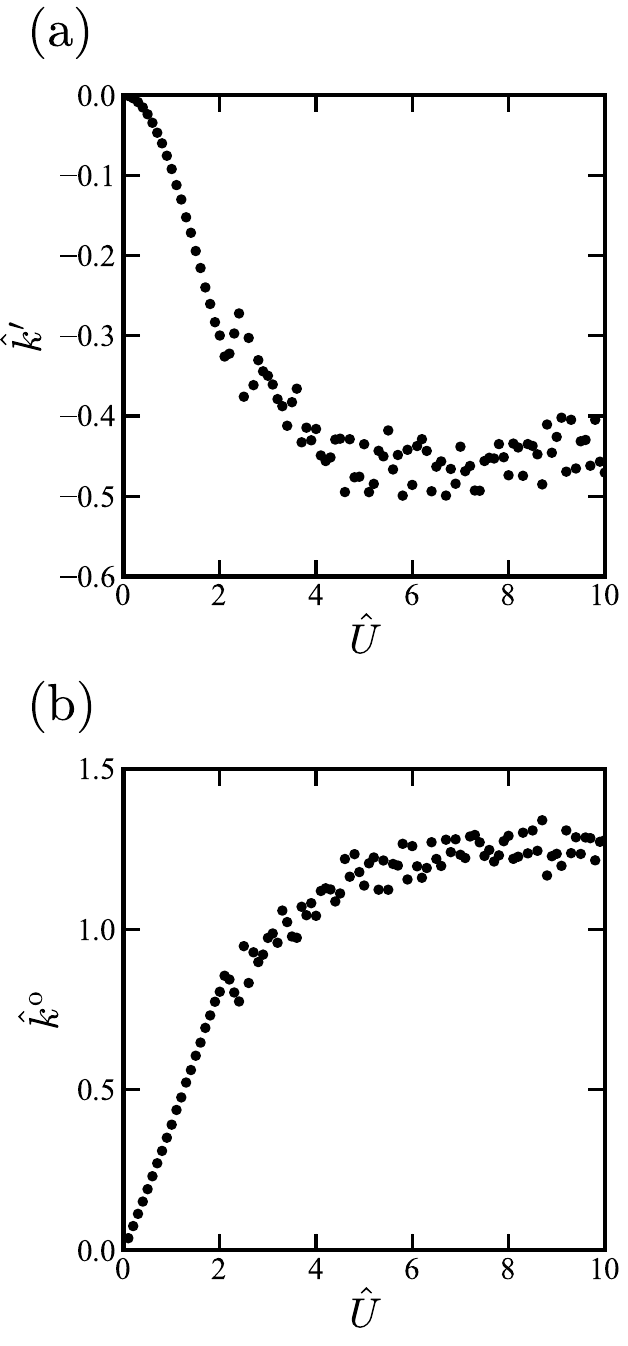}
\caption{
(a) The effective off-diagonal even elasticity $\hat{k}'=k'/K$ for fully trained microswimmers as a function 
of $\hat{U}$ [see Eq.~(\ref{eff_even_offd})].
The $\hat{U}$-dependence is similar to that of the non-reciprocality $\hat{R}$ in Fig.~\ref{performance_plot}(a).
(b) The effective odd elasticity $\hat{k}^{\rm o}=k^{\rm o}/K$ as a function of $\hat{U}$ [see Eq.~(\ref{eff_odd})].
The $\hat{U}$-dependence closely resembles that of the loop frequency $1/\hat{T}$ in Fig.~\ref{performance_plot}(b).
}
\label{odd_ela_plot}
\end{figure}

Finally, we explain the method to extract the effective even and odd elasticities of the trained microswimmer from the 
numerical data.
A direct way is to assume the following odd-elastic Hookean relations between the forces and the spring 
extensions~\cite{Yasuda22,Kobayashi23b,Fruchart2021}:
\begin{align}
    \begin{pmatrix}
        F_{\rm A} \\ F_{\rm B}
    \end{pmatrix}
    = 
    \begin{pmatrix}
        k^{\rm e} & k' + k^{\rm o}
        \\ 
        k'-k^{\rm o} & k^{\rm e}
    \end{pmatrix}
    \begin{pmatrix}
        u_{\rm A} \\ u_{\rm B}
    \end{pmatrix}. 
\label{eq_hookean}
\end{align}
In the above expression, the forces $F_\alpha$ are given by Stokes' law 
$F_\alpha = - 6 \pi \eta a \dot{u}_\alpha$ and $u_{\alpha} = L_\alpha - \ell$ as before.
In the elastic matrix, $k^{\rm e}$ and $k'$ represent the diagonal and off-diagonal even elasticities, 
respectively, while $k^{\rm o}$ represents the effective odd elasticity.
The diagonal even elasticity $k^{\rm e}$ should be distinguished from the spring elastic constant $K$ 
in the elastic microswimmer model.

Typically, a finite $k^{\rm e}$ results in damping oscillation.
Given the fact that our data demonstrate sustained oscillations without amplitude decay, we assume 
in the following that $k^{\rm e}=0$. On the other hand,
the off-diagonal even elasticity $k'$ characterizes the fore-aft amplitude asymmetry as described 
in Fig.~\ref{amplitude}, and the odd elasticity $k^{\rm o}$ should quantify the non-reciprocal dynamics 
of the microswimmer~\cite{Kobayashi23b,Yasuda22}.

\begin{figure*}[tb]
\centering
\includegraphics[scale=0.52]{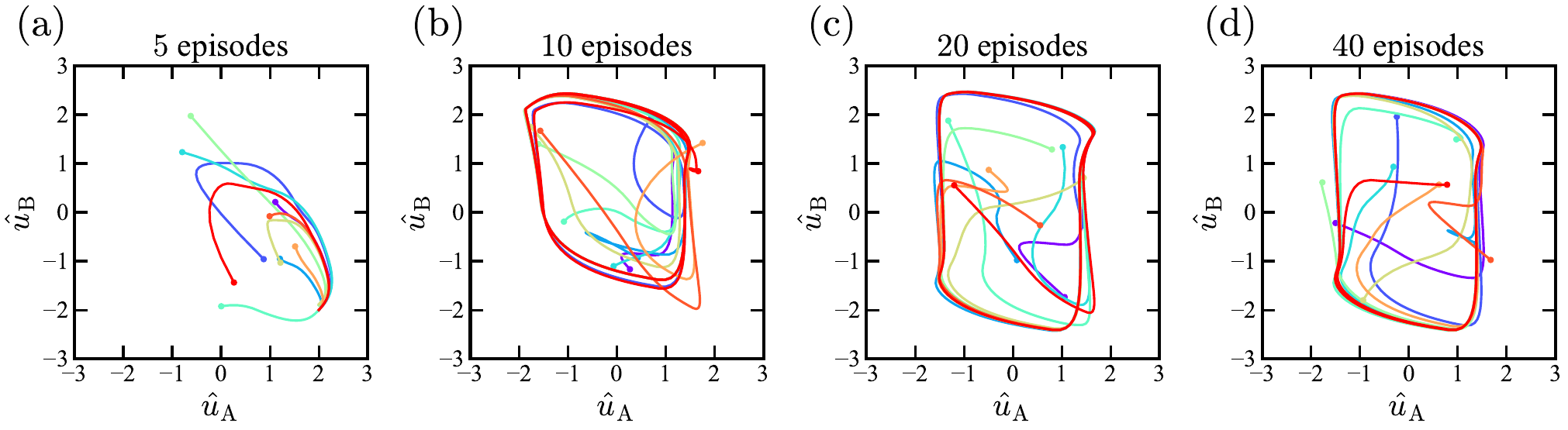}
\caption{
The configuration space trajectories at different training stages; (a) 5, (b)10, (c) 20, and (d) 40 episodes for a 
microswimmer with $\hat{U}=2$. 
Each trajectory is initiated from a distinct initial state, indicated by different colored dots.
Training beyond the stage (d) aims to approach the optimized limit cycle shown in Fig.~\ref{ex_u0_to_u10}(d)
for 200 episodes.
}
\label{training_prigress_plot}
\end{figure*}

To obtain $k'$ and $k^{\rm o}$ from the numerical data, we calculate the cross-correlations 
$\overline{F_{\rm A} u_{\rm B}}$ and $\overline{F_{\rm B} u_{\rm A}}$ averaged over one period of the 
deformation cycle. 
This is possible since the trained dynamics for the two springs are periodic with the same frequency and 
a finite phase difference.
Since these correlations are given by
\begin{align}
\overline{F_{\rm A} u_{\rm B}} &= \left(k' + k^{\rm o}\right)\overline{u^2_{\rm B}},
\\
\overline{F_{\rm B} u_{\rm A}} &= \left(k' - k^{\rm o}\right)\overline{u^2_{\rm A}},
\end{align}
the effective (off-diagonal) even and odd elastic coefficients can be obtained from 
\begin{align}
k' & =\frac{1}{2}\left(\frac{\overline{F_{\rm A} u_{\rm B}}}{\overline{u_{\rm B}^2}}+\frac{\overline{F_{\rm B} u_{\rm A}}}{\overline{u_{\rm A}^2}}\right), 
\label{eff_even_offd}
\\
k^{\rm o}& =\frac{1}{2}\left(\frac{\overline{F_{\rm A} u_{\rm B}}}{\overline{u_{\rm B}^2}}-\frac{\overline{F_{\rm B} u_{\rm A}}}{\overline{u_{\rm A}^2}}\right), 
\label{eff_odd}
\end{align}
where all the related correlations can be directly calculated from the numerical data.

In Figs.~\ref{odd_ela_plot}(a) and (b), we plot dimensionless even elasticity, $\hat{k}'=k'/K$ and 
odd elasticity, $\hat{k}^{\rm o}=k^{\rm o}/K$, respectively, as functions of $\hat{U}$.
Both $\hat{k}'$ and $\hat{k}^{\rm o}$ vanish when $\hat{U}=0$.
We recognize that the behavior of $\hat{k}'$ is similar to that of $\hat{R}$ in Fig.~\ref{performance_plot}(a)
(except a constant shift), while the data of $\hat{k}^{\rm o}$ closely resembles that of $1/\hat{T}$ 
in Fig.~\ref{performance_plot}(b).
These results indicate that $\hat{k}'$ characterizes the extent of amplitude asymmetry that reduces the 
enclosed area, whereas $\hat{k}^{\rm o}$ directly corresponds to the loop frequency.
Since $1/\hat{T} \propto \hat{k}^{\rm o}$, we confirm that the average swimming velocity $\overline{V}$ in 
Eq.~(\ref{dimensionless_speed}) is proportional to the odd elasticity $k^{\rm o}$, i.e., $\overline{V} \propto k^{\rm o}$.
Notably, the odd elasticity $\hat{k}^{\rm o}$ is proportional to the actuation velocity up to $\hat{U} \le 2$,
similar to $1/\hat{T}$.

The fact that the effective odd elasticity $\hat{k}^{\rm o}$ is proportional to the loop frequency 
$1/\hat{T}$ is consistent with our model of a stochastic odd microswimmer in which the 
presence of odd elasticity was implemented~\cite{Yasuda21,Kobayashi23a}. 
In the odd microswimmer model, the probability flux forms a closed loop and 
the eigenvalues of the corresponding frequency matrix are proportional to the odd elasticity.
For the time-correlation functions in general odd Langevin systems~\cite{Yasuda22}, 
it was shown that odd elasticity determines the frequency of the sinusoidal component both in their 
symmetric and anti-symmetric parts.
These results support the relationship wherein the work per cycle is generally determined by the 
product of the odd elasticity and the closed loop area in active systems~\cite{Scheibner2020OddElasticity_ml,Fruchar2023}.

\section{Training progression of elastic microswimmer}
\label{sec:training}

In this section, we discuss briefly the dynamic evolution of swimming behavior during the training process. 
The plots in Figs.~\ref{training_prigress_plot}(a)-(d) illustrate the configuration space 
trajectories $\hat{u}_\alpha$ ($\alpha = \rm{A, B}$) at distinct stages of training in 
Fig.~\ref{u2_plot}(a), namely, 5, 10, 20, and 40 episodes, respectively, when $\hat{U}=2$.
Each trajectory starts from a different initial state denoted by colored dots.

In the initial stage of training, as shown in Fig.~\ref{training_prigress_plot}(a), the swimming behavior is 
relatively restricted.
Starting from different initial states, the system only evolves towards the point $(2, -2)$, showing limited 
exploration of the motion possibilities. 
This early stage primarily focuses on optimizing short-term rewards where cyclic motion has not emerged yet. 
As the microswimmer gains more training experience, its behavior progressively improves. 
After approximately 10 episodes, as shown in Fig.~\ref{training_prigress_plot}(b), the microswimmer begins 
to explore long-term rewards accessible through cyclic motions. 
Enclosed loops emerge in the configuration space, and the microswimmer starts to recognize the dynamics 
required for sustainable cyclic locomotion.

In Fig.~\ref{training_prigress_plot}(c) at around 20 training episodes, the shape of loops becomes 
clearer and more refined. 
This stage shows the swimmer's ability to fine-tune its motion strategy toward the optimal cyclic 
pattern to maximize long-term rewards.
After training for 40 episodes, the microswimmer achieves the optimized limit cycle, as shown 
in Fig.~\ref{training_prigress_plot}(d).
Importantly, this optimized cycle is robust across various initial conditions, and it will eventually 
approach the limit cycle shown in Fig.~\ref{ex_u0_to_u10}(d).

\section{Summary and discussion}
\label{sec:summary}

Using deep reinforcement learning (Deep Q-Network), we have investigated how the effective odd 
elasticity emerges when optimizing the swimming ability of an elastic microswimmer~\cite{Pande15,Pande17,Yasuda17c}.
One of the key findings is the optimized natural-length dynamics without the need for prescribed motion. 
Notably, we observed a strategy transition (the emergence of waiting behavior) when the actuation velocity $\hat{U} \approx 2$ 
(Fig.~\ref{ex_u0_to_u10}). 
For larger $\hat{U}$, the trained microswimmers adapt to the slow hydrodynamic relaxation and avoid the velocity decrease.
This waiting strategy significantly improves the swimming ability compared to the elastic microswimmer 
with prescribed dynamics having large-frequency oscillations.
Additionally, the trained microswimmer exhibits fore-aft asymmetry in the spring amplitudes (Fig.~\ref{amplitude}), 
which is generally difficult to predict and implement in the prescribed dynamics.

By calculating the force-displacement correlations for fully trained microswimmers, we have extracted 
the effective even and odd elasticities, $k'$ and $k^{\rm o}$, respectively (Fig.~\ref{odd_ela_plot}).
We have shown that the $U$-dependencies of $k'$ and $k^{\rm o}$ closely resemble those of the 
non-reciprocality $R$ in Fig.~\ref{performance_plot}(a) and the loop frequency $1/T$ in 
Fig.~\ref{performance_plot}(b), respectively. 
From the numerical data, we have further confirmed that the average swimming velocity $\overline{V}$ 
is proportional to the cycle performance $R/T$ (Figs.~\ref{performance_plot}(c) and (d)), as 
predicted in Eq.~(\ref{dimensionless_speed}).
These results clarify the proportionality between the average velocity and odd elasticity,
$\overline{V} \propto k^{\rm o}$.
Our study demonstrates the use of machine learning to reveal the emergence of odd elasticity in various 
active systems.

If we assume that the energy injected through the non-reciprocal process balances with the 
dissipation due to the net motion of a microswimmer, its power (work per unit time) scales as 
$\dot{W} \sim (\eta a \overline{V}) \times \overline{V} \propto (k^{\rm o})^2$.
Here, $\eta a \overline{V}$ corresponds to the dissipative force, and we have used the relation 
$\overline{V} \propto k^{\rm o}$~\cite{Kobayashi23b}. 
For an odd microswimmer, it was further shown that all the extracted work due to odd elasticity is converted 
into the entropy production rate~\cite{Yasuda21}.
Hence, odd elasticity is useful to characterize non-reciprocal dynamics of active micromachines, 
including not only microswimmers but also other molecular motors~\cite{Yasuda22-machlup,Yasuda22}. 
With our analysis method, the work performance of these micromachines can be quantified 
by observing their deformation, even without any precise understanding of their specific functions.

In our machine learning model, we adopted the DQN algorithm with a discrete action space to train elastic microswimmers. 
Since the swimming ability is directly related to the effective strength of non-reciprocal forces, the optimized dynamics
are to adjust one of the natural lengths to either its maximum or minimum value before changing the other.
For such strategies, the microswimmer does not change the two natural lengths simultaneously, and thus 
continuous-action algorithms are unnecessary.
To validate this argument, we have also employed other continuous-action algorithms, such as Deep Deterministic Policy 
Gradient (DDPG)~\cite{Lillicrap2015} and Soft Actor-Critic (SAC)~\cite{Haarnoja2018}.
Both of these methods resulted in the same strategy as the discrete DQN algorithm. 
Although these continuous-action algorithms are helpful for more complicated microswimmers, 
the discrete DQN algorithm used in this work is sufficient for an elastic microswimmer moving
in a one-dimensional space.

Microswimmers composed of biomaterials are often soft and exhibit viscoelastic responses. 
This softness is crucial not only because of the inevitable interaction with viscoelastic environments but also due 
to their versatile functionalities~\cite{schamel2014,walker2015,wu2018,medina2018,huang2019,Striggow2020,soto2022}, 
such as mechanical signal sensing~\cite{xu2019,huo2020}, cargo loading and unloading~\cite{alapan2028, xu2020}, 
and navigation through intricate channels~\cite{vutukuri2020}. 
On the other hand, when one considers manipulating the gait-switching of a microswimmer from a 
simple potential field, our elastic model becomes more suitable. 
For example, if we consider manipulating the spheres through optical tweezers or a harmonic electromagnetic 
field~\cite{Grosjean2016, huang2019}, the corresponding potential could be determined effectively through 
Eqs.~(\ref{f1})-(\ref{f3}).

We comment that the even and odd elasticities obtained from the numerical data are assumed to be linear.
In a more general scenario, however, these elasticities can be non-linear~\cite{Fruchar2023}. 
Since a diagonal positive elasticity generally leads to a decaying oscillation~\cite{Lin23}, non-linearity 
is commonly required for odd-elastic systems to exhibit a stable limit cycle~\cite{Ishimoto22_ml}.
The inclusion of non-linear elasticity can describe more general cases, such as the spontaneous onset 
of oscillations with a specific amplitude that is regulated by the ratio of linear and non-linear 
even elasticities~\cite{Ishimoto23}. 
In the present work, the oscillation amplitudes are determined by the constraint that clips the 
natural lengths, and hence the linear approach is more suitable.

In our study, a crossover between the elastic- and hydrodynamic-dominated limits generally exists between microswimmers' large- and small-frequency limits. For such systems, we have shown that a swimming transition strategy around the crossover actuation velocity is desired for the optimized performance because a simple prescribed periodic motion becomes less efficient in the large-frequency regime. Crucially, we emphasize a fundamental characteristic that exists in such systems: a significant delay between the applied control ($\ell_\alpha$) and the shape responses ($L_\alpha$). This delay induces strategy transitions that can generally exist in various microswimmers.

Applications that use machine learning to optimize gait-switching for microswimmer navigation have been widely studied in recent literatures~\cite{cichos2020, stark2021, hartl2021, stark2023, alageshan2020, landin2021, schneider2019, nasiri2023, tsang2020, zou2022, qin2023, liu2023, borra2022, zhu2022}.
These studies have provided valuable insights into the potential of machine learning to optimize the dynamics of microswimmers.
In the current work, in addition to focusing on performance optimization, we have emphasized the possibility of extracting useful information such as non-reciprocality and odd elasticity. 
In future studies, we aim to establish universal relations between performance and odd elasticity for 
different types of microswimmers~\cite{Laugabook}.

\begin{acknowledgements}

We thank J.\ Shuai and X.\ Xu for the useful discussion.
L.S.L.\ is supported by Tokyo Human Resources Fund for City Diplomacy.
K.Y.\ acknowledges the support by a Grant-in-Aid for JSPS Fellows (No.\ 22KJ1640) from the 
Japan Society for the Promotion of Science (JSPS).
K.I.\ acknowledges the JSPS, KAKENHI for Transformative Research Areas A (No.\ 21H05309) 
and the Japan Science and Technology Agency (JST), FOREST Grant (No.\ JPMJFR212N).
S.K.\ acknowledges the support by the National Natural Science Foundation of China (Nos.\ 12274098 and 
12250710127) and the startup grant of Wenzhou Institute, University of Chinese Academy of Sciences 
(No.\ WIUCASQD2021041).
L.S.L., K.Y., and K.I.\ were supported by the Research Institute for Mathematical Sciences, an International 
Joint Usage/Research Center located in Kyoto University.
K.Y., K.I., and S.K.\ acknowledge the support by the JSPS Core-to-Core Program ``Advanced core-to-core 
network for the physics of self-organizing active matter” (No.\ JPJSCCA20230002).
\end{acknowledgements}


\end{document}